\documentclass[11pt,a4paper]{article}

\usepackage{amsmath, amssymb, mathrsfs}
\usepackage{braket}
\usepackage{cite}

\usepackage[top=25mm, bottom=25mm, left=25mm, right=25mm]{geometry}
\usepackage{titlesec}
\titlespacing*{\section}{0pt}{1ex}{-0.1ex}
\titlespacing*{\subsection}{0pt}{1ex}{-0.1ex}
\titlespacing*{\subsubsection}{0pt}{0ex}{-0.2ex}

\usepackage[colorlinks=true, linkcolor=blue, urlcolor=blue, citecolor=blue]{hyperref}

\usepackage{tikz}
\usetikzlibrary{decorations.pathmorphing}

\definecolor{internationalorange}{rgb}{1.0, 0.31, 0.0}

\newcommand\RR{\ensuremath{\mathbb R}}

\newcommand\td{\text{d}}
\newcommand\cO{{\cal O}}
\newcommand\cS{{\cal S}}
\newcommand{\skyp}{{\cal I}^+}
\newcommand{\skym}{{\cal I}^-}
\newcommand{\bz}{\bar{z}}
\newcommand{\bw}{\bar{w}}

\newcommand{\p}{\partial}

\newcommand{\be}{\begin{equation}}
\newcommand{\ee}{\end{equation}}
\newcommand{\bea}{\begin{eqnarray}}
\newcommand{\eea}{\end{eqnarray}}

\def \ga {\gamma_{z\bz}}
\def \gai {\gamma_{z\bz}^{-1}}

\def \gawi {\gamma_{w\bw}^{-1}}

\begin{document}
\thispagestyle{empty}
\begin{flushright}
\tt DESY-21-102
\end{flushright}

\vskip 0pt
\begin{center}
{\LARGE\sffamily\bfseries
Infrared photons and asymptotic symmetries
}
\vskip 35pt

{\large\sc Zhengwen Liu}
\\
{\it Deutsches Elektronen-Synchrotron DESY, Notkestrasse 85, 22607 Hamburg, Germany}

\vskip 5pt
{\large\sc Pujian Mao}
\\
{\it
Center for Joint Quantum Studies and Department of Physics,\\
School of Science, Tianjin University, 135 Yaguan Road, Tianjin 300350, China}
\end{center}
\vskip 10mm

\begin{quote}
\textbf{Abstract~}
S-matrix elements exhibit universal factorization when multiple infrared photons are emitted in scattering processes. We explicitly show that the leading soft factorization of tree-level amplitudes with the emission of any number of soft photons can be interpreted as the Ward identity of the asymptotic symmetry of gauge theory.
\end{quote}

\section{Introduction}
Understanding the factorization property of scattering amplitudes in various special kinematics, such as soft, collinear and Regge limits, plays an important role in both making precision predictions for physical observables and revealing the hidden structure of quantum field theory.
In particular, on-shell scattering amplitudes may display universal factorization when one or more soft particles are emitted in scatterings, in e.g.~gauge theories and gravity \cite{Low:1958sn,Weinberg:1964ew,Weinberg:1965nx,Burnett:1967km,Jackiw:1968zza,Gross:1968in,Bassetto:1984ik,Berends:1988zn,Catani:1999ss,DelDuca:1990gz,Cachazo:2014fwa}.
The universal soft factorization of scattering amplitudes is often referred to as {\it soft theorems} in the literature.

In recent years there have been renewed interests on soft theorems from a more theoretical side.
The new enthusiasm comes in part from the discovery of a remarkable connection between soft theorems and symmetries at null infinity \cite{Strominger:2013lka,Strominger:2013jfa}.
The soft theorems can be rewritten as Ward identities of certain {\it asymptotic symmetries}. In particular, the large gauge transformations of electromagnetism and non-Abelian gauge theory are behind the soft photon theorem \cite{He:2014cra} and the soft gluon theorem \cite{He:2015zea,Mao:2017tey}. The constraints on emission of soft gravitons come from the supertranslation \cite{He:2014laa}, the Abelian ideal part of the Bondi-Metzner-Sachs (BMS) transformations \cite{Bondi:1962px,Sachs:1962wk,Sachs:1962zza}. More details on this connection can be found in the review \cite{Strominger:2017zoo}.

While soft theorems in a variety of theories have been intensively investigated from various perspectives, the study of the related asymptotic symmetries has been limited to single and double soft limits. The purpose of this work is to extend the relationship between asymptotic symmetries and soft theorems to any multiplicity for soft particles.
For simplicity, we choose tree-level amplitudes with multiple soft-photon emission as our testing ground in this work.
First, we derive the leading soft theorem for any number of soft photons using Feynman diagrams and the eikonal approximation.
We then extend the mysterious relation between the single soft photon theorem and the large gauge transformations of electromagnetism to any number of soft photons.

\section{The multi-soft photon current}\label{sec-soft-theorem}

In this section we present the soft-current for the emission of any number of soft photons from a tree-level amplitude with an arbitrary number of hard particles, in order to reveal its relation with asymptotic symmetries in subsequent sections.

We consider an on-shell scattering amplitude with any $n$ {\it hard} external legs and $m$ {\it soft} photons ${\cal M}_{m+n}(p_1,\ldots, p_n; q_1,\ldots,q_m)$, where $\{p_1, \ldots, p_n\}$ and $\{q_1, \ldots, q_m\}$ denote hard and soft momenta respectively.
Introducing an infinitesimal parameter $\lambda$, $q_i=\mathcal{O}(\lambda)$ and $p_i=\mathcal{O}(1)$ in the soft limit $\lambda\to 0$.
In the multi-soft limit, any on-shell amplitude is expected to factorizes into a product of a universal operator acting on the on-shell amplitude with only hard particles \cite{Berends:1988zn}
\begin{align}\label{soft-leading}
{\cal M}_{m+n}(p_1,\ldots, p_n; q_1,\ldots,q_m) \,=\, \mathscr{J}_m(q_1,\ldots,q_m)\,{\cal M}_{n}(p_1,\ldots, p_n) + \cdots,
\end{align}
where the ellipsis denotes terms that are power-suppressed in the soft limit.
It is convenient to define the soft current with the polarization tensor structure stripped off
\begin{align}\label{soft-current}
 \mathscr{J}_m(q_1,\ldots,q_m) \,\equiv\,
 \varepsilon^{\mu_1}(q_1)\cdots \varepsilon^{\mu_m}(q_m)
 \mathcal{J}_{\mu_1\cdots\mu_m}(q_1,\ldots,q_m),
\end{align}
where $\varepsilon^{\mu_i}(q_i)$ is the polarization vector of the soft photon $i$, we assume which satisfies $q_i \cdot \varepsilon(q_i) = 0$.

It is well-known that the current for the emission of a single soft photon takes the form in the all-outgoing convention \cite{Low:1958sn,Weinberg:1964ew},
\begin{align}\label{single-soft-current}
\mathcal{J}^{\mu}(q) \,=\, \sum_{i=1}^{n} e_i\, {\cal E}_i^{\mu}(q),
\qquad
{\cal E}_{i}^\mu(q) \,\equiv\, {p_i^\mu \over p_i\cdot q},
\end{align}
where $e_i$ denotes the charge of the hard particle $i$, and ${\cal E}_i^{\mu}(q)$ is the famous {\it eikonal vertex}.
It is straightforward to verify that the eikonal current is conserved, i.e.~$q^\mu\mathcal{J}_{\mu}(q) = 0$, due to the conservation of charge.

It is clear that the leading soft current receives the contributions from only diagrams where soft particles are emitted from external hard (charged) lines.
To derive the multi-soft photon current, let us introduce a generalization of the eikonal vertex.
We define a $m$-point {\it soft effective vertex} as a sum of all possible contributions for the emission of $m$ soft photons from a single hard external line.
Schematically,
\begin{align}\label{}
\begin{aligned}
\begin{tikzpicture}[scale=1]
 \draw[line width=1.0pt] (-1.2, 0) -- (1.2,0);
 \draw[line width=0.6pt, decorate, decoration={snake, amplitude=1.1pt, segment length=5pt}, color=internationalorange] (128:1.5) -- (0,0);
 \fill[] (130:1.5) circle (0) node[above] {$1$};
 \draw[line width=0.6pt, decorate, decoration={snake, amplitude=1.1pt, segment length=5pt}, color=internationalorange] (102:1.5) -- (0,0);
 \fill[] (102:1.5) circle (0) node[above] {$2$};
 \draw[line width=0.6pt, decorate, decoration={snake, amplitude=1.1pt, segment length=5pt}, color=internationalorange] (52:1.5) -- (0,0);
 \fill[] (50:1.5) circle (0) node[above] {$m$};
 \fill[] (0,0) circle (2.7pt);
 \fill[color=internationalorange] (85:1.3) circle (0.8pt);
 \fill[color=internationalorange] (77:1.3) circle (0.8pt);
 \fill[color=internationalorange] (69:1.3) circle (0.8pt);
 \fill[] (2,0) circle (0) node[] {\Large $=$};
 \draw[xshift=42mm, line width=1.0pt] (-1.5, 0) -- (1.5, 0);
 \draw[xshift=42mm, xshift=-8mm, decorate, decoration={snake, amplitude=1.1pt, segment length=5pt}, line width=0.6pt, color=internationalorange] (0, 1.3) -- (0,0);
 \fill[xshift=42mm, xshift=-8mm,] (0, 1.3) circle (0) node[above] {$1$};
 \draw[xshift=42mm, xshift=-2mm, decorate, decoration={snake, amplitude=1.1pt, segment length=5pt}, line width=0.6pt, color=internationalorange] (0, 1.3) -- (0,0);
 \fill[xshift=42mm, xshift=-2mm,] (0, 1.3) circle (0) node[above] {$2$};
 \draw[xshift=42mm, xshift=8mm,decorate, decoration={snake, amplitude=1.1pt, segment length=5pt}, line width=0.6pt, color=internationalorange] (0, 1.3) -- (0,0);
 \fill[xshift=42mm, xshift=8mm,] (0, 1.3) circle (0) node[above] {$m$};
 \fill[xshift=42mm, ] (-0.8,0) circle (1.3pt);
 \fill[xshift=42mm, ] (-0.2,0) circle (1.3pt);
 \fill[xshift=42mm, ] (0.8,0) circle (1.3pt);
 \fill[xshift=42mm, color=internationalorange] (0.3,1.1) circle (0.8pt);
 \fill[xshift=42mm, color=internationalorange] (0.1,1.1) circle (0.8pt);
 \fill[xshift=42mm, color=internationalorange] (0.5,1.1) circle (0.8pt);
 \fill[] (6.0,0.2) circle (0) node[right] {\large $+~\operatorname{permutation}(1,\ldots,m).$};
\end{tikzpicture}
\end{aligned}
\end{align}
Evidently, the single soft effective vertex is just the eikonal vertex \eqref{single-soft-current}
\begin{align}\label{}
{\cal J}_{i}^\mu(q) \,=\, e_i\, {\cal E}_{i}^\mu(q).
\end{align}
There are two different configurations for the emission of two soft partons from a single line.
To be precise, the double soft effective vertex is given by
\begin{align}\label{}
{\cal J}_{i}^{\mu_1\mu_2}(q_1, q_2)
\,&=\, e_i^2\, \Big({\cal E}_{i}^{\mu_2}(q_{12})\, {\cal E}_{i}^{\mu_1}(q_{1}) + {\cal E}_{i}^{\mu_1}(q_{12})\, {\cal E}_{i}^{\mu_2}(q_{2}) \Big)
\nonumber\\
\,&=\, e_i^2\, {\cal E}_{i}^{\mu_1}(q_{1})\, {\cal E}_{i}^{\mu_2}(q_{2})
\\
\,&=\, {\cal J}_{i}^{\mu_1}(q_{1})\, {\cal J}_{i}^{\mu_2}(q_{2}),
\nonumber
\end{align}
with $q_{a\cdots b} = q_a + \cdots + q_b$.
Similarly, there are $3!=6$ diagrams that contribute to the soft effective vertex when three soft photons are emitted from a hard line.
Therefore the triple soft effective vertex is given by
\begin{align}\label{}
{\cal J}_{i}^{\mu_1\mu_2\mu_3}(q_1, q_2, q_3)
\,&=\, e_i^3\, \Big({\cal E}_{i}^{\mu_3}(q_{123})\, {\cal E}_{i}^{\mu_2}(q_{12}) \, {\cal E}_{i}^{\mu_1}(q_{1})+ \text{permutations} \Big)
\nonumber\\
\,&=\, e_i^3\, {\cal E}_{i}^{\mu_1}(q_{1})\, {\cal E}_{i}^{\mu_2}(q_{2})\,{\cal E}_{i}^{\mu_3}(q_{3})
\\
\,&=\, {\cal J}_{i}^{\mu_1}(q_{1})\, {\cal J}_{i}^{\mu_2}(q_{2}) \, {\cal J}_{i}^{\mu_3}(q_{3}).
\nonumber
\end{align}
As briefly outlined in \cite{Weinberg:1965nx}, by induction it is easy to obtain the soft effective vertex for arbitrary $m$ soft photons
\begin{align}\label{soft-effective-vertex-m}
{\cal J}_{i}^{\mu_1\cdots \mu_m}(q_1, \ldots, q_m)
\,&=\, {\cal J}_{i}^{\mu_1}(q_1)\, \cdots \, {\cal J}_{i}^{\mu_m}(q_m).
\end{align}
Very nicely, the multi-soft photon effective vertex in \eqref{soft-effective-vertex-m} has a fully factorized form, i.e.~a $m$-point soft effective vertex factorizes into a product of $m$ eikonal vertices.
As we will see below, soft effective vertices defined above not only are useful building blocks to construct
full soft currents, but also capture the key structure of the full soft current \cite{Liu:2020softPartonNote}.

In the following, we show how to use soft effective vertices to derive the full soft current.
The single-soft current is simply the summation of eikonal vertices over all hard legs,
\begin{align}\label{soft-current-1}
\mathcal{J}^{\mu}(q) \,=\, \sum_{i=1}^{n} {\cal J}_i^{\mu}(q).
\end{align}
To compute the double-soft photon current, we need to consider all tree-level Feynman diagrams involving two photons and $n$ external hard lines that carry momenta $p_i$ and charges $e_i$.
There are two types of diagrams that contributes to the result:~two soft photons emit from one and two hard lines respectively
\begin{align}\label{}
\mathcal{J}^{\mu_1\mu_2}(q_1,q_2) \,&=\,
\Bigg( \sum^n_{i,j=1 \atop i\ne j} \mathcal{J}_i^{\mu_1}(q_1) \mathcal{J}_j^{\mu_2}(q_2) \Bigg)
 + \Bigg( \sum^n_{i=1} \mathcal{J}_i^{\mu_1\mu_2}(q_1,q_2) \Bigg)
\nonumber \\
 \,&=\,
\Bigg( \sum^n_{i=1} \mathcal{J}_i^{\mu_1}(q_1) \Bigg) \Bigg( \sum^n_{j=1} \mathcal{J}_j^{\mu_2}(q_2) \Bigg)
\\
\,&=\,\mathcal{J}^{\mu_1}(q_1)\, \mathcal{J}^{\mu_2}(q_2).
\nonumber
\end{align}
More generally, for the multi-soft current, the $m$ soft photons can emit from at most $m$ external hard lines.
Using the property that any $m$-point soft photon effective vertex factorizes into a product of $m$ eikonal vertices, i.e.\,\eqref{soft-effective-vertex-m}, a sum over all possible diagrams gives
\begin{align}\label{}
\mathcal{J}^{\mu_1\cdots\mu_m}(q_1,\ldots,q_m) \,=\, \mathcal{J}^{\mu_1}(q_1)\,\cdots\, \mathcal{J}^{\mu_m}(q_m).
\end{align}
As expected, the soft-current for the emission of any number of soft photons takes a nicely factorized form.

\section{Soft theorem as Ward identity of asymptotic symmetries}
In this section, we show how to recover the soft photon theorems with any number of soft photons presented in the previous section as Ward identities of large gauge transformation. We begin by reviewing the derivation of the single soft photon theorem from large gauge transformation in \cite{He:2014cra, Conde:2016csj}. In this section, we restrict ourself to the case of massless hard particles.

\subsection{Single-soft photon theorem}
The Minkowski spacetime has two null boundaries, past null infinity $\skym$, and future null infinity $\skyp$. Here we will concentrate on $\skyp$, while everything can be repeated on $\skym$ in a similar way. The retarded spherical coordinates will be applied with the following change of coordinates:
\begin{equation}
\label{retard}
u=t-r\,,\quad x^1+ix^2=\frac{2rz}{1+z\bz}\,,\quad x^3=r\,\frac{1-z\bz}{1+z\bz}\,,
\end{equation}
where $r=\sqrt{x^ix_i}$. The line element of Minkowski spacetime now becomes
\begin{equation}
\label{metric}
\td s^2=-\td u^2-2\td u\td r+2r^2\gamma_{z\bz}\td z\td\bz\,,\quad \gamma_{z\bz}=\frac{2}{(1+z\bz)^2}\,.
\end{equation}
The piece $2\gamma_{z\bz}\td z\td\bz$ is just the metric of the round sphere $S^2$. $\skyp$ is the submanifold $r\rightarrow\infty$, with topology $S^2\times\RR$. We will denote the sphere at $u=\pm\infty$ by $\skyp_\pm$.

The soft momentum $q$ can be characterized by an energy $\omega_q$ and a direction on the sphere $(w,\bw)$, as
\begin{equation}
\label{qmu}
 q_{\mu}=\omega_q\left(1,\frac{w+\bar{w}}{1+w\bar{w}},i\frac{\bar{w}-w}{1+w\bar{w}},\frac{1-w\bar{w}}{1+w\bar{w}}\right)\ .
\end{equation}
For massless hard particles, the momenta $p_k$ can be parameterized as~\eqref{qmu} by energies $\omega_k$ and directions $(w_k,\bw_k)$ in a similar way. The single soft photon theorem~\eqref{single-soft-current} can be then rewritten in the position space as
\begin{gather}
\label{soft0+}
	\lim_{\omega_q\to0}\langle\textrm{out}|\omega_q\,\mathfrak{a}_{+(-)}(q)\cS|\textrm{in}\rangle=
	\frac{1+|w|^2}{\sqrt{2}}\sum_{k=1}^n \frac{e_k}{w-w_k}\,\langle\textrm{out}|\cS|\textrm{in}\rangle\ ,
\end{gather}
where $e_k$ is the electric charge of the $k$-th particle, and $\mathfrak{a}_{+(-)}(q)$ is the annihilation operator that creates outgoing negative(positive)-helicity soft photon with momentum $q$.

To connect with asymptotic symmetries, we recall that if a symmetry is generated by a charge $Q$, the associated Ward identity reads as follows:
\begin{equation}
\label{Ward}
	\langle\rm{out}|Q^{\textrm{out}}\cS-\cS Q^{\textrm{in}}|\rm{in}\rangle=0 \ ,
\end{equation}
which is a consequence of $[Q,\cS]=0$. The charge for a spontaneously broken symmetry must act non-linearly on the states, otherwise it would annihilate the vacuum. So we can decompose the charge into linear and non-linear pieces $Q=Q_{\rm{L}}+Q_{\rm{NL}}$. The Ward identity for the charge of a broken symmetry becomes
\begin{equation}
\label{WardNL}
	\langle\rm{out}|Q_{\rm{NL}}^{\textrm{out}}\cS - \cS Q^{\textrm{in}}_{\rm{NL}}|\rm{in}\rangle=
	-\langle\rm{out}|Q^{\textrm{out}}_{\rm{L}}\cS - \cS Q^{\textrm{in}}_{\rm{L}}|\rm{in}\rangle\,.
\end{equation}
Since $Q_{\textrm{NL}}$ creates zero-momentum Goldstone boson, equation~\eqref{WardNL} is very similar to the single soft photon theorem~\eqref{soft0+}.

The charge responsible for the single soft photon theorem is given by \cite{Ashtekar:1987tt,Barnich:2001jy,He:2014cra}
\begin{equation}
\label{Q0}
 Q_{\varepsilon_\textrm{out}}=-\int_{\skyp_{-}} \td z \td\bz \,\ga\, \varepsilon(z,\bz)\, A_u^0 \ ,
\end{equation}
where $\varepsilon(z,\bz)$ is an arbitrary function on the sphere that generates the large (residual) gauge transformations of the form $\delta A_{\mu}=\partial_{\mu}\varepsilon(z,\bz)$ in the radial gauge condition $A_r=0$. It is a symmetry of the S-matrix following \cite{Strominger:2013lka,He:2014cra}. The gauge field and the conserved current that coupled to the gauge field are assumed to follow a $\frac1r$-expansion as
\begin{equation}
A_u=\frac{A^0_u(u,z,\bz)}{r}+\cO\Big(\frac{1}{r^2}\Big) \ ,\quad A_{z(\bz)}=A^0_{z(\bz)}(u,z,\bz) + \sum\limits_{m=1}^\infty \frac{A^m_{z(\bz)}(u,z,\bz)}{r^m} \ ,
\end{equation}
\begin{equation}
J_u=\frac{J^0_u(u,z,\bz)}{r^2}+\cO\left(\frac{1}{r^3}\right) \ , \quad 	J_{z(\bz)}=\frac{J^0_{z(\bz)}(u,z,\bz)}{r^2} + \sum\limits_{m=1}^\infty \frac{J^m_{z(\bz)}(u,z,\bz)}{r^{m+2}} \ .
\end{equation}
We have used the ambiguities of a conserved current to set the radial component of the current to zero. This is consistent with working in the radial gauge. Maxwell's equations yield
\be
e\,\p_u A^0_u =\gai \p_u(\p_z A^0_{\bz} + \p_{\bz} A^0_z) + J^0_u \ .
\ee
The charge $Q$ splits into
\begin{align}
\label{Q0NL}
	Q_{\textrm{NL}}=\int_{\skyp}\td z\td\bz \td u\,\varepsilon\,\partial_u\partial_{\bz}A^0_z \ ,\\
\label{Q0L}
	Q_{\textrm{L}}=\frac12\int_{\skyp}\td z\td\bz \td u\,\gamma_{z\bz}\,\varepsilon\,J_u^0 \,.
\end{align}
For notational brevity, we will suppress the \textit{out} label and keep only the anti-holomorphic term.

For the gauge field, we perform a stationary-phase approximation of the mode expansion:
\begin{equation}
	A^0_{z(\bz)}(x)=-\frac{i}{8\pi^2}\frac{\sqrt{2}}{1+z\bz}\int_0^\infty\td \omega_q\left[
\mathfrak{a}_{+(-)}(\omega_q\hat{x})\,e^{-i\omega_q u}-\mathfrak{a}^\dagger_{-(+)}(\omega_q\hat{x})\,e^{i\omega_q u}\right] \ ,
\end{equation}
where the creation and annihilation operators satisfy the standard commutation relations. Then, using the Fourier relation (defining $F(u)=\int_{-\infty}^{\infty}\td\omega\,e^{i\omega u}\tilde{F}(\omega)$):
\begin{equation}
	\int_{-\infty}^{\infty}\td u\,\partial_uF(u)=2\pi i\lim_{\omega\to0}\left[\omega\tilde{F}(\omega)\right] \ ,
\end{equation}
and the special choice
\begin{equation}
\label{eps}
	\varepsilon(z,\bz)=\frac{1}{w-z} \ ,
\end{equation}
we obtain for the non-linear piece of the charge\footnote{
We only keep the anti-holomorphic part of the charge~\eqref{Q0NL}, namely the charge only contains $\partial_{\bz}\varepsilon$. In particular one needs to split~\eqref{Q0L} via $J^0_u\to\frac12J^0_u+\frac12J^0_u$. Otherwise an extra factor of 2 that arise from a proper treatment of the radiative phase space~\cite{He:2014cra,Mohd:2014oja} is needed.
}:
\begin{align}
\label{Q0NLbk}
	\langle\textrm{out}|Q_{\textrm{NL}}\cS|\textrm{in}\rangle&=\frac{1}{4}\frac{\sqrt{2}}{1+|w|^2}\lim_{\omega_q\to0}
	\langle\textrm{out}|\omega_q\,\mathfrak{a}_+(q)\cS|\textrm{in}\rangle \ ,
\end{align}
For the linear piece, considering the complex scalar charged (with charge $Q_e$) matter as example where the current at leading order is $J^0_\mu=i Q_e(\bar{\Phi}^0\p_\mu \Phi^0 - \Phi^0\p_\mu \bar{\Phi}^0)$, we just need to use the boundary canonical commutation relation \cite{Lysov:2014csa}:
\begin{equation}
 [\bar{\Phi}^0(u,z,\bz),\Phi^0(u',w,\bw)]=\frac{i}{4}\gawi\, \Theta(u'-u)\delta^2(z-w) \ ,
\end{equation}
to obtain that
\begin{align}
\label{Q0Lbk}
 \langle\textrm{out}|Q_{\textrm{L}}\cS|\textrm{in}\rangle&=\sum_{k=1}^n
 -\frac{e_k}{4(w-w_k)}\langle\textrm{out}|\cS|\textrm{in}\rangle \ .
\end{align}
Assembling the two expressions, the soft theorem~\eqref{soft0+} can be recovered from~\eqref{WardNL}.

We have only paid attention to the \textit{out} part of~\eqref{WardNL}. The analysis of the \textit{in} part can be carried out in the same way. The generators on $\skyp$ and $\skym$ are connected by an anti-podal identification, see, e.g., \cite{He:2014cra} for details.

\subsection{Double-soft photon theorem}
After having the connection between the single soft photon theorem and the asymptotic symmetries, it is natural to ask about the case involving more soft photons. Let us start with the double soft case, for which we need to consider a family of Ward identity
\be
\langle\textrm{out}|[Q_1,[Q_2,\cS]]|\textrm{in}\rangle=0 \ .
\ee
Since $[Q_2,\cS]=0$, the crucial part for deriving the connection between the Ward identity and double soft photon theorem is
\be
[Q_{1NL},[Q_{2},\cS]]=0 \ ,
\ee
which is equivalent to
\be
[Q_{1NL},[Q_{2NL},\cS]]=-[Q_{1NL},[Q_{2L},\cS]] \ .
\ee
Similar to the gravity case \cite{H:2018ktv}, this expression can be rewritten using the Jacobi identity among $Q_{1NL}$, $Q_{2L}$ and $\cS$ as
\be\label{double}
[Q_{1NL},[Q_{2NL},\cS]]=[Q_{2L},[\cS,Q_{1NL}]] + [\cS,[Q_{1NL},Q_{2L}]] \ .
\ee
The bracket structure should be considered as the large gauge transformation by the first generator on the second one. Since there is no gauge field in the expression of the linear piece of the charge~\eqref{Q0L}, it is gauge invariant and does not generate any gauge transformation. So the linear piece of the charge commutes with the non-linear piece of the charge, i.e.,
\be\label{commutator}
[Q_{1NL},Q_{2L}]=0 \ .
\ee
Substituting this back to~\eqref{double} and using the fact that $[\cS,Q_{1}]=0$, one can derive
\be
[Q_{1NL},[Q_{2NL},\cS]]=(-1)^2[Q_{2L},[Q_{1L},\cS]] \ .
\ee
Using the last expression and repeating the derivation in the previous subsection by inserting twice the charges, one can recover the double soft photon theorem in a straightforward way.

\subsection{Triple-soft photon theorem}
The derivation in the previous subsection can be easily extended to triple soft photon case where the crucial ingredient is
\be
[Q_{1NL},[Q_{2NL},[Q_3,\cS]]]=0 \ .
\ee
After splitting the third charge into linear and non-linear pieces, one can obtain
\be\label{triple}
[Q_{1NL},[Q_{2NL},[Q_{3NL},\cS]]]= -[Q_{1NL},[Q_{2NL},[Q_{3L},\cS]]] \ .
\ee
First, repeating the derivation in the double soft case for $Q_{2NL}$, $Q_{3L}$, and $\cS$, the Ward identity \eqref{triple} can be rewritten as
\be
[Q_{1NL},[Q_{2NL},[Q_{3NL},\cS]]]= [Q_{1NL},[Q_{3L},[Q_{2L},\cS]]] \ .
\ee
Then, using the Jacobi identity among $Q_{1NL}$, $Q_{3L}$ and $[Q_{2L},\cS]$ and the commutation relation between $Q_{1NL}$ and $Q_{3L}$, we obtain
\be
[Q_{1NL},[Q_{2NL},[Q_{3NL},\cS]]]= [Q_{3L},[Q_{1NL},[Q_{2L},\cS]]] \ .
\ee
Finally, repeating the derivation in the double soft case for $Q_{1NL}$, $Q_{2L}$, and $\cS$, we get
\be
[Q_{1NL},[Q_{2NL},[Q_{3NL},\cS]]]= (-1)^3 [Q_{3L},[Q_{2L},[Q_{1L},\cS]]] \ .
\ee
This is the key ingredient for recovering the triple soft photon theorem from the Ward identity of asymptotic symmetry.

\subsection{Multi-soft photon theorem}

In this subsection, we will show that the multi-soft photon theorem can be derived from the Ward identity
\be\label{n}
[Q_{1NL},[Q_{2NL},......,[Q_{(m-1)NL},[Q_{m},\cS]......]=0 \ .
\ee
We will prove it by induction. Suppose that the $(m{-}1)$-soft photon theorem can be obtained from the Ward identity
\be
[Q_{1NL},[Q_{2NL},......,[Q_{(m-2)NL},[Q_{m-1},\cS]......]=0 \ ,
\ee
which means that the Ward identity can be rewritten as
\be
[Q_{1NL},[Q_{2NL},......,[Q_{(m-1)NL},\cS]......]=(-1)^{m-1} [Q_{(m-1)L},[Q_{(m-2)L},......,[Q_{1L},\cS]......] \ .
\ee
Implementing the relation $[Q_{(m-1)NL},\cS]=-[Q_{(m-1)L},\cS]$, one obtains
\be\label{useful}
(-1)^{m} [Q_{(m-1)L},......,[Q_{1L},\cS]......]=[Q_{1NL},......,[Q_{(m-2)NL},[Q_{(m-1)L},\cS]......] \ .
\ee
For the case of $m$ soft photons, after splitting the $m$th charge into the linear and non-linear pieces, the Ward identity becomes
\be
[Q_{1NL},......,[Q_{mNL},\cS]......]=-[Q_{1NL},......,[Q_{(m-1)NL},[Q_{mL},\cS]......] \ .
\ee
First, applying the relation in \eqref{useful} for $Q_{2NL}$,......,$Q_{(m-1)NL}$, $Q_{mL}$ and $\cS$, the Ward identity can be rewritten as
\be
[Q_{1NL},[Q_{2NL},......,[Q_{mNL},\cS]......]=(-)^{m-1}[Q_{1NL},[Q_{mL},[Q_{(m-1)L},......,[Q_{2L},\cS]......] \ .
\ee
Then, using the Jacobi identity among $Q_{1NL}$, $Q_{mL}$ and $[Q_{(m-1)L},......,[Q_{2L},\cS]......]$ and the commutation relation between $Q_{1NL}$ and $Q_{mL}$, we obtain
\be
[Q_{1NL},[Q_{2NL},......,[Q_{mNL},\cS]......]=(-)^{m-1}[Q_{mL},[Q_{1NL},[Q_{(m-1)L},......,[Q_{2L},\cS]......] \ .
\ee
Repeating such procedure for $m-2$ times to put the $Q_{1NL}$ to the right, we get
\be
[Q_{1NL},[Q_{2NL},......,[Q_{mNL},\cS]......]=(-)^{m-1}[Q_{mL},[Q_{(m-1)L},......,[Q_{2L},[Q_{1NL},\cS]......] \ .
\ee
Finally, using $[Q_1,\cS]=0$, we obtain
\be
[Q_{1NL},......,[Q_{mNL},\cS]......]=(-1)^m [Q_{mL},......,[Q_{1L},\cS]......] \ .
\ee
This is precisely what we need to recover the multiple soft photon theorem from the Ward identity \eqref{n} of asymptotic symmetry.

\section{Conclusions and Discussions}

To conclude, we have generalized the equivalence between the single-soft photon theorem and the Ward identities of asymptotic symmetries to any number of soft photons. Remarkably, the Ward identity for the $m$-photon theorem can be derived from the Ward identity for the single soft photon theorem by induction. The derivation of the $m$-photon Ward identity simply involves the Jacobi identity and the commutator of asymptotic symmetry generators. This unexpected simplification fits very well with the linearity of the theory. Soft photons can be inserted independently into the amplitudes and any number of soft photons insertion corresponds any number of independent insertion of the conserve currents.

Let us comment on several directions for future research.
First, our analysis in this work was restricted to leading soft theorems.
It would be interesting to continue our exploration beyond leading order in the multiple soft limit.
Second, it would be interesting to investigate the relations between multi-soft theorems and asymptotic symmetries in non-Abelian gauge theory and gravity.
Third, it has been observed that the single soft photon theorem is equivalent to the electromagnetic memory effect that is a residual velocity to
the test charges in the detector after the passage of electromagnetic radiation \cite{Pasterski:2015zua}. It is natural to ask for the memory effect connected to the multi-soft limit of amplitude.

Finally, while the study of asymptotic symmetries until now has been restricted to theoretical interests, their use to study scattering amplitudes relevant to real-life collider processes has somewhat been limited.
It would be very tempting to see if these novel ideas provide insight into the structure of the infrared singularities of amplitudes in gauge theory.
This would help us improve theoretical predictions for multi-parton processes at the Large Hadron Collider.

\section*{Acknowledgments}

ZL would like to thank Claude Duhr for collaboration on related topics.
We are grateful to Volker Schomerus and Jun-Bao Wu for comments and a careful reading of the manuscript.
The work of ZL was supported by the Deutsche Forschungsgemeinschaft (DFG) under Germany's Excellence Strategy (EXC 2121) `Quantum Universe' (No.\,390833306), and by the European Research Council (ERC) under the European Union’s Horizon 2020 research and innovation programme through the ERC Consolidator Grant ``{\it Precision Gravity:~From the LHC to LISA}'' (No.\,817791). The work of PM is supported in part by the National Natural Science Foundation of China under Grants No.\,11905156 and No.\,11935009.

\providecommand{\href}[2]{#2}\begingroup\raggedright\endgroup

\end{document}